\documentclass[pra,showpacs,amsmath,a4paper,twocolumn]{revtex4}
\usepackage{geometry}
\usepackage{bbm}
\usepackage{amssymb}
\usepackage{amsmath}
\usepackage{graphicx}
\usepackage{amsfonts}
\usepackage{hyperref}
\usepackage{oldgerm}
\usepackage{color}
\usepackage{epsfig}
\usepackage{here}
\usepackage{enumerate}

\newcommand{\be}{\begin{eqnarray}}
\newcommand{\ee}{\end{eqnarray}}
\newcommand{\ket}[1]{|#1\rangle}

\begin{document}
\title{Weak randomness in device independent quantum key distribution and the advantage of using high dimensional entanglement}
\author{Marcus Huber$^{1,2,3}$, Marcin Paw{\l}owski$^{1,4}$}

\affiliation{$^1$University of Bristol, Department of Mathematics, Bristol, BS8 1TW, U.K.}
\affiliation{$^2$ICFO-Institut de Ciencies Fotoniques, 08860 Castelldefels, Barcelona, Spain}
\affiliation{$^3$Universitat Autonoma de Barcelona, 08193 Bellaterra, Barcelona, Spain}
\affiliation{$^4$Instytut Fizyki Teoretycznej i Astrofizyki, Uniwersytet Gdanski, PL-80-952 Gdansk, Poland}

\begin{abstract}

We show that in device independent quantum key distribution protocols the privacy of randomness is of crucial importance. For sublinear test sample sizes even the slightest guessing probability by an eavesdropper will completely compromise security. We show that a combined attack exploiting test sample and measurement choices compromises the security even with a linear size test sample and otherwise device independent security considerations. We explicitly derive the sample size needed to retrieve security as a function of the randomness quality. We demonstrate that exploiting features of genuinely higher dimensional systems one can reduce this weakness and provide device independent security more robust against weak randomness sources.
\end{abstract}


\maketitle
\section{Introduction}
Quantum cryptography is one of the most prominent applications of quantum information theory. It enables a level of security in key distribution that is unparalleled in classical information theory, as a successful eavesdropping would violate fundamental laws of quantum physics. This is true at least in theoretically perfect settings. Specific attacks targeted at imperfect technical implementations, such as e.g. low efficiency detectors have shown that there is still a lot of room for improvement until the application is perfected. The original proposals however were still making a lot of strong assumptions on the systems used for cryptography. One can group these approaches in two classes. Prepare and measure quantum key distribution (QKD) utilizes the fact that measurement is not possible without disturbance in quantum physics (the first protocol was proposed in Ref.~\cite{BB84}). The strong assumption here is that one is in perfect control of the source and measurement apparatuses. Indeed it was shown that security can be compromised if the source of the information carriers, or the measurement apparatuses used for interacting with them is manipulated \cite{QKDbreaking1,QKDbreaking2}. In the second type of protocols entanglement is used to establish a secure key, which due to the limited shareability of quantum correlations provides security even if the source of entangled states is in the hands of an eavesdropper (the first such protocol was proposed in Ref.~\cite{Ekert91}). Also in this case however it is possible to break the security of the protocols if the eavesdropper also has manipulated the measurement devices. Fortunately, using device independent verification of entanglement, one can overcome this flaw and recent works have focused on device independent quantum key distribution (DIQKD) (see e.g. Refs.~\cite{DIQKD00,DIQKD01,DIQKD1,DIQKD2}).\\
All device independent proposals so far have used assumptions about a perfectly uniform randomness being readily available. In Ref.~\cite{Honza} it was shown that even a slight imperfection in randomness generation leads to a possible loophole and even entanglement based protocols can be compromised. This loophole however originates in the sublinear size of the test sample.

The attack of Eve in that paper assumes that Eve is responsible for the weak randomness and can use her knowledge of the bias to guess with high probability in which rounds the security check will be performed and thus remain undetected. However, this is only one of the points where the randomness enters the protocol. The other is the choice of the measurement settings. As we will see the bad quality of randomness used there has a big impact on the security. In this paper we generalize the attack from Ref.~\cite{Honza} and show that the security of the DIQKD can be compromised, even when using a linear test sample, if Eve exploits the min-entropy loss in both the choice of the settings and the test sample.

We show that below a certain threshold of randomness quality key generation is no longer possible with qubit protocols. Furthermore we propose a scheme that overcomes this weakness by considering genuinely high-dimensional entangled systems, that are readily available in quantum photonics (see e.g. Refs.~\cite{dimexp1,dimexp2,dimexp3,dimexp4,dimexp5}).

This paper is structured as follows: First we present the scenario that we are working in and the protocol that the parties are using. Then we derive the minimal violation of the Bell inequality used as a security parameter as a function of min-entropy loss rate. Next we prove the necessity of a feature that any QKD protocol must have in presence of weak randomness: a linear size of the test sample. Then we find the sufficient size of the sample for CGLMP \cite{CGLMP} testing. We end with discussing the implications of our work and the open problems.

\section{The Setting}
To start let us first specify the setting. Alice and Bob want to share a secure key. They implement a protocol under the following conditions
\begin{enumerate}
\item Potentially compromised measurement apparatus. Alice and Bob have access to quantum measurement devices, which they cannot trust. However, following \cite{DIQKD3}, we assume that the observables measured in the different runs commute.
\item Potentially compromised source of multidimensionally entangled quantum systems.
\item No pre-existing secret key: This is very important. If Alice and Bob have some shared secret bits they could use it as a randomness source with perfect randomness.
\item Weak randomness: Alice and Bob have potentially compromised sources of randomness. The min-entropy loss rate of their is $L$. We assume that the randomness generators of Alice and Bob can be correlated.
\item An authenticated classical channel.
\end{enumerate}

Now we specify the protocol that they are going to implement. It is the standard DIQKD setting. Two communicating parties are going to use the CGLMP \cite{CGLMP} inequality violation as their security parameter. 
To estimate the violation each party has to randomly choose one of the settings $a=0,1$ for Alice and $b=0,1$ for Bob. To generate the key Bob will use a third setting $b=2$ which gives him outcomes maximally correlated with Alice's when she chooses $a=0$. The protocol has $N$ runs. First Bob randomly chooses a subset of $fN$ of runs where the parties estimate the parameter. In this subset he chooses settings $b=0$ or 1 randomly. In the remaining $(1-f)N$ runs he uses $b=2$. Alice chooses $a=0$ or 1 randomly in all the runs.

After the measurements for all the runs are complete Bob announces the cases which he used for parameter estimation. The parties announce the settings and outcomes for all of these runs and use them to estimate the value of the CGLMP inequality violation. In the remaining $(1-f)N$ runs Alice announces when she has chosen $a=1$ and the parties discard these runs. Only the cases when $a=0$ and $b=2$ are used to generate the key.

This protocol is quite standard for DIQKD apart from the fact that usually only a number of runs sublinear in $N$ is used for the parameter estimation. The reason why we need a sample of linear size is the weak randomness in the possession of the parties. From \cite{Honza} we already know that Alice and Bob cannot have any secure protocol with a smaller test sample under these conditions (and we give the explicit proof for the DIQKD scenario in the Appendix section (\ref{ap3})).

In order to analyze the protocol we take the same approach as Ref.\cite{Honza} in quantifying the imperfection of the randomness using the min-entropy loss rate. Min-entropy is the measure of choice as any source generating the bits that has any type of randomness will also exhibit non-zero min-entropy. This measure makes our result completely general as it is quite easy to compute the min-entropy loss rate of any source specified using a different measure. E.g. Santha-Vazirani (S-V) sources \cite{SV} have a min-entropy loss rate of $L=1+\log(\frac{1}{2}-\epsilon)$. Therefore, our result can in a straightforward manner be applied to S-V sources. The same holds for sources which are bit-fixing, biased or described by some Renyi entropy \cite{book}. On the other hand, if we would choose any other measure of randomness relating it to others would be not possible, e.g. bit fixing sources, which simply fix some fraction of the bits in advance and choose the rest at random, are indistinguishable from a fully deterministic one from an S-V point of view.

Let $(M,b)$ denote an imperfect source of randomness that creates strings of length $M$, according to a probability distribution with min-entropy $b$. We quantify the bias of the source by the min-entropy loss rate denoted $L = \frac{M-b}{M}$.

\section{The most general attack exploiting bad randomness in the sifted key generation}

The protocol described above generates sifted key which can be later turned into the secret key via classical privacy amplification and error correction procedures. Though these procedures also require randomness their analysis falls outside the scope of this paper and we are interested in the most general attack on the "quantum" part of the protocol. First let us focus on the randomness in the choice of the settings.

The min-entropy loss rate $L$ is the resource that the adversary uses to attack the protocol. It can be directly related to her probability of guessing the settings in each round of the protocol. For clarity, we can divide the guessing in two parts. The first is deciding whether $b=2$ or, in other words, whether this round is used for parameter estimation. The second is guessing the measurement settings in each round.

The goal of Eve is to learn as much of the sifted key as possible while remaining undetected. When $b=2$ then the adversary aims at maximizing her correlations with Alice. If $b<2$ her aim is to hide her interference. The limits of her resource make it impossible to know the value of $b$ in every round. Since the strategy optimal for $b=2$ gives her more information about the key than the one for $b<2$ the optimal attack is to use the strategy for $b=2$ even in some rounds used for parameter estimation provided that she can avoid detection.

The strategy optimal for $b=2$ is to prepare a product state (its details are discussed later) but if $L$ is large enough we will see that it becomes also the optimal strategy for $b<2$. In this case we have no hopes for security.

Our protocol has two important parameters: the amount of CGLMP violation and the fraction $f$ of the rounds used for parameter estimation. Now we find the lower bounds for both of these as the functions of $L$.

\section{Bell inequality violation as a security check}

In the device independent protocols the key ingredient is the parameter estimation phase where the parties estimate the violation of the Bell inequality. However to test it some randomness is required. To our knowledge, in all the works on device independent protocols it is assumed that this randomness is perfect. The attack presented in \cite{Honza} used only the fact that weak randomness can let Eve to choose a subset of runs where she knows that they are not used for parameter estimation. But weak randomness of the settings leads to the increase of the local bound which, in turn, leads to another loophole in quantum cryptography \cite{Caslav}. Though, apart from \cite{Caslav} there have been other works that studied the dependence of the local bound on the input randomness, they have been either restricted to CHSH \cite{Noah} or the randomness was measured in the terms of conditional Shannon entropy \cite{M-niid2}. Therefore, we need to adapt the methods from \cite{M-niid2} to find the local bound on CGLMP as a function of the min-entropy loss rate. \\
We start by expressing CGLMP in a "normalized" form
\begin{align}
\label{cglmp}
\frac{1}{4}\left(P(A\leq B|a=0,b=0)+P(B\leq A|a=0,b=1)+\right.\nonumber\\
\left.P(A\leq B|a=1,b=1)+P( B<A|a=1,b=0)\right) \leq \frac{3}{4},
\end{align}
where $a$ and $b$ denote the settings of Alice and Bob respectively and $A$ and $B$ their outcomes. It is a normalized variant of CGLMP first introduced in \cite{ZG}.

\begin{figure}[th!]
\begin{center}
  \includegraphics[scale=0.8]{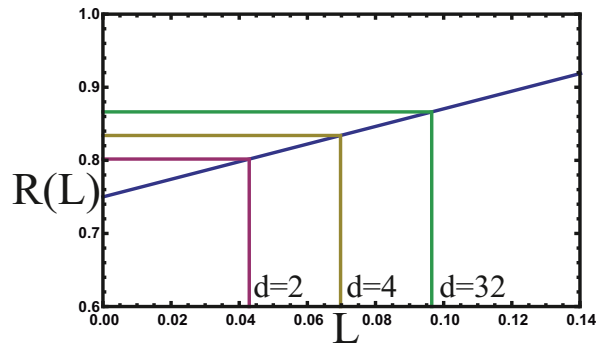}
  \caption{(Color online) Here we plot the maximal expectation value of the normalized CGLMP inequality ,with product states that are distributed by Eve, as a function of the min-entropy loss rate $L$. For comparative purposes we also include three exemplary violations by maximally entangled states of dimension $d=2$, $d=4$ and $d=32$. It follows directly from this relation that if the min entropy loss rate $L$ exceeds $\approx 0.043$, there is no hope of a secure protocol using qubits.}{\label{RL}}
\end{center}
\end{figure}

One approach to Bell inequalities is to treat them as nonlocal games. We can think of the parties receiving their inputs from the referee who assures them that they are chosen according to uniform probability distribution. He can, however, be wrong or lying. What happens then is Alice and Bob playing the game with the strategy optimized for the uniform distribution of settings while they are not. Effectively, they are trying to violate inequality
\begin{align}
\label{cglmp2}
p_{00}P(A\leq B|a=0,b=0)+p_{01}P(B\leq A|a=0,b=1)+\nonumber\\
p_{10}P(A\leq B|a=1,b=1)+p_{11}P( B<A|a=1,b=0) \nonumber\\
\leq R,
\end{align}
where $p_{ij}$ is the probability of Alice getting setting $i$ and Bob $j$. Furthermore, these probabilities change each round. In the Appendix section (\ref{ap1}) we give the details of the proof that the optimal violation that can be achieved with product states for a given min entropy loss rate $L$ is given by
\be
R(L)=\frac{3}{4}+\frac{L}{2(2-\log 3)}.
\ee
This bound is plotted on Fig \ref{RL} and it becomes the crucial local bound that Alice and Bob need to violate if they want to have a chance for security.

\section{Security considerations}

However violating the local bound is only a necessary condition for the security. It is known \cite{DIQKD3} that the key rate in the DIQKD protocol with commuting observables secure against general attacks is lowerbounded by
\be
K\geq H_{\min}(A|E)-\frac{N_{pub}}{N_{key}},
\ee
where $H_{\min}(A|E)$ is the min-entropy rate of the Alice's bit of key conditioned on Eve's information, $N_{pub}$ the amount of communication in the error correction and privacy amplification phases and $N_{key}$ the length of the key. In the Appendix sections (\ref{ap4},\ref{ap2}) we show that the violation of the local bound in CGLMP inequality implies $H_{\min}(A|E)>0$ and indeed we can even infer a lower bound $H_{\infty}(A|E)\geq  \frac{R_{obs}-R(L)}{2\ln2}$ where $R_{obs}$ is the average value of the CGLMP measured by the parties.

Error correction and privacy amplification imply that whenever $a=0$ and $b=2$ the correlations are provably perfect. This forces Eve to use the states that give maximal correlation between the parties for settings $a=0$ and $b=2$ as the part of her strategy optimal for $b=2$. It is in contrast with the attack proposed in \cite{Honza} where in most of the rounds not used for parameter estimation Eve tries to decrease the correlations between Alice and Bob. This attack can remain undetected under the assumption of sublinear sample size.

If the violation of CGLMP is large enough we know that we can have a secure protocol but we still have to find out the size of the test sample $f$.

From the considerations presented earlier we know that the optimal strategy for Eve is to use the strategy optimal for $b<2$ in $kN$ rounds with $k\leq f$ and the strategy optimal for $b=2$ in the rest of them. The strategy optimal for $b<2$ is, obviously, to send the state that violates CGLMP the most for the given number of outcomes. The strategy optimal for $b=2$ is to send the product state $\ket{\psi}$. Of course the closer Eve wants to bring $k$ to $f$, the more min-entropy loss $L$ she has to induce
\be
L_N =\frac{\log{N \choose f N}-\log{(1-k)N \choose (1-f) N}}{\log{N \choose f N}},
\ee
which for large $N$ approaches
\begin{align} \label{L}
L(k,f)=\lim_{N\to \infty}L_N=\nonumber\\
\frac{-f \log(f) - (1-k) \log (1-k) +( f-k) \log( f-k)}{h(f)},
\end{align}
where $h(.)$ is Shannon's binary entropy function.

At the same time the closer $k$ is to $f$ the larger is the Bell inequality violation that the parties can observe. In $\frac{k}{f}$ of the rounds used to estimate CGLMP the state that violates it maximally for the given number of outcomes is distributed. Let us denote this violation  by $R_Q(L,d)$. In the remaining $\frac{f-k}{k}$ rounds the state distributed is $\ket{\psi}$ and, since it is a product state, the maximal violation is $R(L)$. Therefore the violation observed can be at most
\be
R_{obs}\leq \frac{k}{f}R_Q(L,d)+\frac{f-k}{f}R(L),
\ee
which implies
\be
k\geq f\frac{R_{obs}-R(L)}{R_Q(L,d)-R(L)}.
\ee

$R_Q(L,d)$ is the maximal quantum violation of CGLMP inequality with $d$ outcomes and the min-entropy loss rate of the randomness of the settings $L$. There are no known methods of finding this value. However we can always bound it by the algebraic bound: $R_Q(L,d)\leq 1$, which implies
\be
k\geq f\frac{R_{obs}-R(L)}{1-R(L)}=k(R_{obs},L).
\ee

Plugging this into (\ref{L}) we obtain
\be
L\geq L(k(R_{obs},L),f),
\ee
which can be solved for any value of $L$ giving a lower and an upper bound on the fraction $f$ of the rounds used for parameter estimation.
\begin{figure}[th!]
\begin{center}
  \includegraphics[scale=0.8]{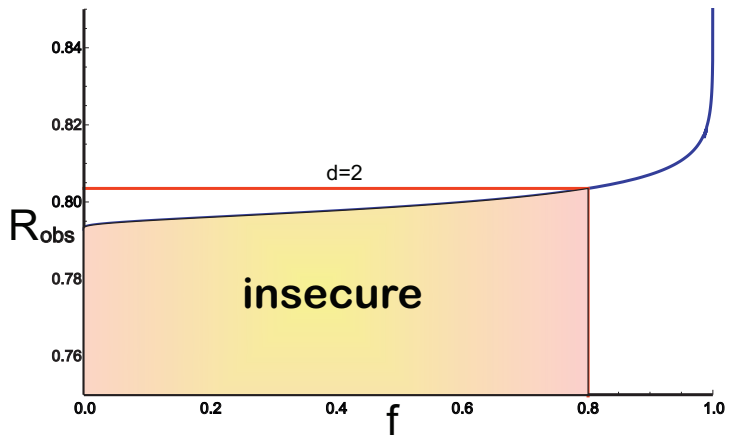}
  \caption{(Color online) Here we depict the lower bound on the observed violation $R_{obs}$ for different linear fractions $f$ and for a fixed value of min-entropy loss rate $L=0.03$. As in this case $R(L)=0.786141$ we know we need to observe a strictly larger violation to provide security, and we see that if the violation is high enough we can indeed choose any $f$. For qubits $d=2$ the fundamental limit is $R_{obs}\leq 0.801777$, which fundamentally constrains $f\leq 0.723026$, whereas with $d=4$ one can already reach $f\leq 0.999416$.}{\label{Robs}}
\end{center}
\end{figure}
There is an upper bound due to the fact that $f$ is the fraction of rounds to be tested with the CGLMP inequality. Here Eve needs to make sure that she restricts most of them to a sample space, that she knows with certainty. Counterintuitively it can happen that if $f$ is chosen too large, that this task actually gets easier for Eve, now succeeding with her attack.\\
Now we also need to discuss another lower bound for $f$. For $N\rightarrow\infty$ this it will also be arbitrarily small due to infinite precision in the measurement of $R_{obs}$. But in every practical scenario we have to face the fact that we only have a limited number of runs and $R_{obs}$ will always carry experimental error bars from its statistical deduction. So depending on the statistical fidelity which we aim for in a finite number of runs, we will have to choose a large enough fraction of $N$ runs to guarantee that such a precision is possible to attain in $fN$ runs. This lower bound is easy to calculate for any specific implementation, but as it depends also on the experimental settings and errors we do not present a detailed analysis of this lower bound. We want to point out, however, that it can in a realistic setting be significantly larger than the upper bound on $f$ for qubits.

\section{Conclusion and Outlook}
In conclusion we have shown that even in DIQKD settings security can be compromised if the local randomness is not uniform. We explicitly derive the fundamental bounds on the randomness quality in this setting, and consequently demonstrate that high-dimensional entanglement is indeed more powerful than mere qubit entanglement. Thus we have not only shed light on the role of randomness in the security of DIQKD protocols but at the same time we have provided a clear example where the generation of high-order entanglement opens up new paths in quantum cryptography.\\
For the fundamental introduction of the protocol we have looked at an idealized scenario. We can however also apply the same reasoning without assuming perfect correlations (discussed in the Appendix section \ref{ap5}) and even deal with attacks exploiting weak randomness in the error correction and privacy amplification phases (see Appendix section \ref{ap6}) or also considering different types of entropy loss in the randomness generators. The basis for the improvement of the protocol remains the same however. A higher violation of Bell-inequalities through higher dimensions increases the randomness of the outcomes and strengthens the protocols, while a higher number of outcomes increases the key rate. A full analysis of different scenarios is under preparation.\\
It should also be noted that this protocol is not necessarily only usable in situations where the randomness quality exceeds a bound for qubits. If the min-entropy loss rate is assumed small enough that the protocol would potentially also be achievable using qubit systems it still pays off greatly to use the high order entanglement, due to the fact that with every measurement multiple bits of key are generated (i.e. $\log(d)$ bits) and there is no downside once such high-dimensional entanglement is readily available. That is proven by numerous recent experimental results that achieve high-order entanglement up to $d\approx 50$ in Ref.\cite{dimexp5}, which provides more than five bits per measurement outcome. Also there exist different possibilities for implementing such high dimensions in photonic systems, e.g. in path (Ref.~\cite{dimexp1}) or a large family of orbital angular momentum (OAM) modes (e.g. Ref~\cite{dimexp2}).\\
\textit{Acknowledgments} We want to thank Sven Ramelow and the members of IQOQI Vienna for initial fruitful discussions. Furthermore we would like to thank Jan Bouda, Martin Plesch, Matej Pivoluska and Andreas Winter for insightful comments on the manuscript. MH also acknowledges funding from the FP7-MarieCurie grant "Quacocos". MP acknowledges support from UK EPSRC, FNP TEAM, ERC grant QUOLAPS and NCN grant No. 2012/05/E/ST2/02352.

\begin{widetext}
\section{Appendix}
\subsection{Impossibility of sublinear sample size}\label{ap3}

In the protocol we use a linear size of the test sample for both CGLMP correlations and the ones for $a=0$ and $b=2$. One could ask if this is indeed necessary. Because the conditions on Eve's attack and the setting of our protocol are substantially different from the one presented in \cite{Honza} we cannot use the attack presented there to prove the insecurity of the protocol. However we can find an attack which works, with slight modifications in both settings. Let us start with the device independent one.

If the test sample is of the size $N^{1-\alpha}$ then Eve in can choose a number $k$ such that $kN>N^{1-\alpha}$. Because we are interested in the limit of large $N$'s $k$ can be chosen arbitrarily small. She exploits weak randomness of the parties to make sure that the $N^{1-\alpha}$ rounds for CGLMP inequality testing are taken from $kN$ predetermined rounds. In these rounds Eve sends the state that Alice and Bob hope to have in all of them. In the rest of the rounds she prepares product state $\ket{\psi}=\ket{x}_{a=0}\ket{x}_{b=2}$ where the indexes denote the bases.

In this case the CGLMP inequality violation is estimated only in the rounds where the state is entangled and, at the same time, the correlations for $a=0$ and $b=2$ are perfect in all the cases. This means that the parties will not detect Eve, while $H_{\infty}(A|E)<k$ can be made arbitrarily small for large $N$.

The min-entropy loss rate in this case can also be made arbitrarily small since
\be
\lim_{N\to\infty}\frac{\log{N \choose  N^{1-\alpha}}-\log{k N \choose  N^{1-\alpha}}}{\log{N \choose N^{1-\alpha}}}=0.
\ee

The version of this attack for the prepare and measure scenario involves Eve not interacting in $kN$ rounds used to check the correlations and measuring the system in $(1-k)N$ rounds in a random basis and sending the state that she got to Bob. It also leads to arbitrarily low min-entropy with arbitrarily good randomness.\\

\subsection{Bell inequality violation as a security check}\label{ap1}

From \cite{M-niid2} we know that the local bound for the CGLMP game played with imperfect randomness is $1-r$ where $r=\min_{a,b}P(a,b)$. The lowest amount of min-entropy for a particular value of $r$ is attained by the distribution $\left(r, \frac{1-r}{3}, \frac{1-r}{3}, \frac{1-r}{3} \right)$. This leads to the local bound in the terms of min-entropy
\be \label{r1}
R\leq\min\left\{3*2^{-H_{\infty}(a,b)},1\right\}
\ee
As soon as this value reaches the quantum bound there is no possibility of the experimental verification that the measured state is not classical and no hopes for secure QKD. The quantum bound, however, depends on the dimension of the measurement system and approaches 1 as $d\to \infty$ \cite{ZG2}. This value is obtained by min-entropy of $\log 3\approx 1.585$, however the critical value is smaller for any state of finite dimension. The quantum bound on the normalized CGLMP inequality (eq.(1) in the main paper) is $0.8177$ for $d=2$ and 0.8516 for $d=5$ \cite{ZG}, which translates to the critical min entropies of $1.875$ and $1.817$ respectively.

For the experiment repeated many times, in the $i$-th run the bound is $R_i=\min\left\{3*2^{-H_{\infty}^i},1\right\}$, where $H_{\infty}^i$ is the min-entropy of the settings in the $i$-th round conditioned on the events from the setting generation for the previous rounds. Let us see how big the average $R=\frac{1}{M}\sum_{i=1}^M R_i$ can be for a given sum of the entropies $H_{\Sigma}=\sum_{i=1}^M H_{\infty}^i$. Clearly, it is pointless for the adversary to set $H_{\infty}^i$ lower than $\log 3$ of any $i$ as it does not increase the bound. In region $H_{\infty}\in [\log 3,2]$ $R$ is convex so the optimal strategy is to use $m$ instances of settings with entropy $\log 3$ and $M-m$ instances with entropy $2$ where
\be
\log3 m+2(M-m)=H_{\Sigma},
\ee
which gives $m=\frac{2M-H_{\Sigma}}{2-\log 3}$
This will give the local bound
\be
R\leq\frac{1}{M}\left(m+\frac{3}{4}(M-m) \right)=\frac{3}{4}+\frac{1-\frac{H_{\Sigma}}{2M}}{2(2-\log 3)}.
\ee
Because the total min-entropy of the source is
\be
H_{\infty}=-\log \max_{a_1,...,a_M,b_1,...,b_M}P(a_1,...,a_M,b_1,...,b_M)
\\
=-\log \max_{a_1,...,a_M,b_1,...,b_M}\prod_{i=1}^M P(a_i,b_i|a_{i-1},...,a_M,b_{i-1},...,b_M)
\\
\leq -\log \prod_{i=1}^M \max_{a_i,b_i}P(a_i,b_i|a_{i-1},...,a_M,b_{i-1},...,b_M)=H_{\Sigma}
\ee
and
\be
L=\frac{2M-H_{\infty}}{2M}=1-\frac{H_{\infty}}{2M}
\ee
we get
\be \label{rl}
R\leq \frac{3}{4}+\frac{L}{2(2-\log 3)}=R(L).
\ee
\subsection{The sufficiency of the violation of the CGLMP inequality}\label{ap4}

{\bf Lemma 1:} The violation of the local bound in CGLMP inequality implies $H_{\min}(A|E)>0$.

{\it Proof.} Let us assume that there exists a setting of Alice, say $a=0$ such that $\max_A P(A|a=0)=1$ which corresponds to zero min-entropy. In such a case instead of measuring with the setting $a=0$ Alice can always just return the outcome which is certain without making any measurement at all. In other words, she can measure observable $\openone$. In \cite{ns-mon} it was shown that if all the observables for one party are compatible (can be measured simultaneously) than no-signaling, quantum and local bounds are the same. But $\openone$ and any observable that is measured for $a=1$ are compatible, so the local bound cannot be violated even by no-signalling theory. QED.
\subsection{Lowerbounding min-entropy of Eve}\label{ap2}

{\bf Theorem} In the scenario presented above $H_{\infty}(A|E)\geq  \frac{R_{obs}-R(L)}{2\ln2}$ where $R_{obs}$ is the average value of the CGLMP measured by the parties.

{\it Proof} Let us take a setting of Alice, say $a=0$ such that $\max_A P(A|a=0)=p$ and consider a procedure similar to the one in lemma but with the outcome of the identity measurement set to the $A$ for which this maximum is reached. From lemma we know that in this case the bound $R$ cannot be violated. However, it still could be violated if for $a=0$ measurement $A_0$ different than $\openone$ is used.

If the value of CGLMP for the strategy with $A_0$ is
\begin{align}
p_{00}P(A\leq B|a=0,b=0)+p_{01}P(B\leq A|a=0,b=1)+\nonumber\\
p_{10}P(A\leq B|a=1,b=1)+p_{11}P( B<A|a=1,b=0)=Q
\end{align}
then the value for the strategy with $\openone$ is at least
\begin{align}
p\big(p_{00}P(A\leq B|a=0,b=0)+p_{01}P(B\leq A|a=0,b=1)\big)+\nonumber\\
p_{10}P(A\leq B|a=1,b=1)+p_{11}P( B<A|a=1,b=0) =\\
Q-(1-p)\big(p_{00}P(A\leq B|a=0,b=0)+p_{01}P(B\leq A|a=0,b=1)\big)
\end{align}
and this has to be lower or equal $R$. Because the outcome probabilities are bounded by 1 and the setting probabilities by exponent of min-entropy we get
\be
Q\leq R+(1-p)2^{1-H_{\infty}}
\ee
or
\be
p\leq 1-2^{H_{\infty}-1}(Q-R).\label{ts}
\ee
In an experiment repeated $M$ times the min-entropy rate of Eve is
\be
H_{\infty}(A|E)=-\frac{1}{M}\log\prod_{i=1}^{M}p_i=-\frac{1}{M}\sum_{i=1}^M\log p_i
\\
\geq\frac{1}{M2\ln2}\sum_{i=1}^M 2^{H_{\infty}^i}(Q_i-R_i)\geq \frac{1}{M2\ln2}\sum_{i=1}^M Q_i-R_i
\\ \label{HL}
\geq \frac{R_{obs}-R(L)}{2\ln2}=H(L).
\ee
Where in the last three inequalities we have used respectively: logarithm's power series expansion, positivity of min-entropy and formula (\ref{rl}). Bound $H(L)$ is far from optimal since the approximations made are pretty coarse. For a specific outcome alphabet much better bounds probably exist.\\
\subsection{Nonperfect Correlations}\label{ap5}
In case one does not assume perfect correlations of the measurement outcomes in the rounds where Alice uses basis $0$ and Bob $2$, due to noise in the system, the situation becomes a little more involved. The basic strategy of Alice, as well as the improvement from higher dimensions remains the same.\\
In this case Eve's resource is still $L$. If it is larger than $0$ it means that some choices of the rounds where the correlations are tested are more probable than the others. Her choice of strategy for each is a pair of numbers the guessing probability $p$ and the Bell expression expectation value $I$. They are connected by a relation $p_G\leq f(I)$. It follows directly from (\ref{ts}) that
\be \label{rel}
p\leq f(I)= 1-2^{H_{\infty}-1}(I-R(L)).
\ee
These numbers for each round have to be chosen in advance and this choice depends on the probability distribution of the tested rounds. Whatever the strategy Eve chooses there is a choice of the test sample which is optimal, from Eve's point of view, for this strategy and for each choice of test sample there is an optimal strategy.

There are two factors that Eve has to consider while choosing her strategy: The observed Bell inequality violation and her average guessing probability of the bit of the key. Eve's target is to maximize the latter while keeping the former above a certain threshold.
If $L<1$ then there is some uncertainty in the choice of the test sample and Eve cannot be sure that the choice will be optimal for her strategy. For every strategy we can list all the choices of the sample according to the value of Eve's target function in the decreasing order. Her guessing probability is the weighted average of them and the only constraint on the weights (which are the probabilities of choosing them as a sample) is that the largest is $p_{max}2^{H_{\infty}}=2^{M(1-L)}$, where $M$ is the amount of bits necessary for the description of the test sample. This means that the best distribution of the probabilities is $(M,p_{max})$-flat.

Because the strategy of Eve is product, i.e. the numbers $I$ and $p_G$ are chosen in advance for each round and do not depend on the actual numbers produced by the generators, it is optimal for her to concentrate her knowledge of the choice of the sample on the information about particular rounds rather than on the relations between them. In other words, it is better for her to know that the sample will be surely tested in round 1 and with probability $50\%$ in round 2 rather than knowing that it will be surely tested in exactly one of these rounds; though the min-entropy is the same in both cases.

Therefore the most general strategy of Eve is to know a fraction $a$ of all the rounds when the Bell inequality is surely tested and a fraction $b$ when it is surely not. If the tested fraction is $f$ then, obviously, $a\leq f$ and $b\leq 1-f$.

The min-entropy loss is
\be
L=\frac{\log{N \choose fN}-\log{(1-a-b)N \choose (f-a)N }}{\log{N \choose fN}},
\ee
observed Bell inequality violation
\be
R_{obs}\leq\frac{a}{f}\beta_{QM}(L)+\frac{f-a}{f}I
\ee
and Eve's average guessing probability
\be
p_{av}=\frac{b}{1-f}+\frac{1-f-b}{1-f}p_G.
\ee
$\beta_{QM}$ is the maximal quantum violation of CGLMP with min-entropy loss rate $L$. It comes from the rounds when Eve is sure that the testing takes place. In the rounds when she knows nothing she uses $p_G$ and $I$ related by (\ref{rel}). These equations can again be numerically solved for any given pair of $L$ and $R_{obs}$ to optimize the strategy and calculate the lower and upper bounds for security considerations. The basic mechanism behind the advantage of higher dimensional systems however remains the same. We plan to extensively survey these generalized scenarios, also including other types of randomness loss in future publications.

\subsection{Error Correction and Privacy Amplification}\label{ap6}

The final stages of every key distribution protocol are Error Correction (EC) and Privacy Amplification (PA). Their purpose is to provide Alice and Bob with a key which copies are identical for both parties and completely unknown to Eve. These stages too require randomness and so far in all the works it has been assumed that this randomness is perfect. If weak randomness is used it opens another avenue of eavesdropper's attacks. In the main text we mention this possibility and state that, since EC and PA are purely classical procedures, they lie out of the scope of this paper. However, in forthcoming work we plan to include them in our analysis and prove the following conjecture:

{\bf Conjecture 1 } - {\it In the scenario discussed in this paper EC and PA are possible if}
\be
H(L)-H(A|B)>L,
\ee
{\it where $H(L)$ is given by formula (\ref{HL}), $H(A|B)$ is the conditional Shannon entropy rate of Alice's key conditioned on Bob's and $L$ the min-entropy loss rate of their source.}

\end{widetext}
\end{document}